\documentclass[aps,prb,twocolumn,groupedaddress,showpacs, nobibnotes]{revtex4-1}
\usepackage{graphicx}
\usepackage{subfigure}
\usepackage{vmargin}
\setpapersize{USletter}
\setmarginsrb{0.8in}{.6in}{0.8in}{0.8in}{0in}{0.2in}{0in}{0in}
\begin{document}

\title{Two-dimensional quantum percolation on anisotropic lattices}
\author{Brianna S. Dillon Thomas}\email{bthomas2@gustavus.edu}
\thanks{\\Current Address: Physics Department, Gustavus Adolphus College, 800 College Ave, St Peter, MN 56082}
\author{Hisao Nakanishi}\email{hisao@purdue.edu}
\affiliation{Department of Physics, Purdue University, West Lafayette, IN 47907}
\date{\today}

\begin{abstract}
In a previous work [Dillon and Nakanishi, Eur. Phys.J B {\bf 87}, 286 (2014)], we calculated 
the transmission coefficient of the two-dimensional quantum percolation model and found there 
to be three regimes, namely, exponentially localized, power-law localized, and delocalized. 
However, the existence of these phase transitions remains controversial, with many other works 
divided between those which claim that quantum percolation in 2D is always localized, and those 
which assert there is a transition to a less localized or delocalized state. It stood out that 
many works based on highly anisotropic two-dimensional strips fall in the first group, whereas 
our previous work and most others in the second group were based on an isotropic square geometry. 
To better understand the difference in our results and those based on strip geometry, we apply 
our direct calculation of the transmission coefficient to highly anisotropic strips of varying 
widths at three energies and a wide range of dilutions. We find that the localization length of
the strips does not converge at low dilution as the strip width increases toward the isotropic 
limit, indicating the presence of a delocalized state for small disorder. We additionally 
calculate the inverse participation ratio of the lattices and find that it too signals 
a phase transition from delocalized to localized near the same dilutions.

\end{abstract}

\maketitle

\section{Introduction}

Quantum percolation (QP) is one of several models used to study transport in disordered systems. 
It is an extension of classical percolation, in which sites/bonds are removed with some 
probability $q$. In the classical percolation model, transmission on a disordered lattice is 
dependent solely on the connectivity of the lattice: the lattice is transmitting only if there 
exists some connected path across the lattice.  As the dilution increases, the probability of 
finding a connected path between any two arbitrarily chosen points decreases while there is 
still a 100\% guarantee of finding {\it some} connected path through the lattice; at some 
critical dilution ($q \approx 41\%$ in site percolation and $q=50\%$ in bond percolation, on the 
square lattice) there is zero chance of finding any connected path across the lattice
and the system is insulating. In the quantum percolation model, the particle traveling through 
the lattice is a quantum one, and thus quantum mechanical interference also influences 
transmission; in fact, even in a perfectly connected lattice ($q=0$) one may have partial 
transmission depending on the energy of the particle and the boundary conditions of the lattice. 
Because of these interference effects, localization will occur at a lower dilution than in the 
classical model, if at all.

The question of whether disorder prevents conduction, especially in two dimensions, is one which 
spans decades, beginning with the introduction of the Anderson model in 1958.\cite{anderson:1958} 
A few decades later, Abrahams et al used one-parameter scaling theory to show that in the 
Anderson model, any amount of disorder prevents conduction for $d \leq 2$. \cite{abrahams:1979} 
This result was confirmed through various studies in the years following, with some exceptions 
for special cases (see for example Refs \onlinecite{lee:1981} and \onlinecite{eilmes:2001} and 
references therein).

Due to its similarities with the Anderson model, it was initially believed that the quantum 
percolation model would likewise lack a transition to a delocalized state in two dimensions. 
However, this proved to not necessarily be the case, and some controversy arose over what 
phase transitions may exist for the 2D QP model. Some found there to be only localized 
states \cite{soukoulis:1991, mookerjee:1995, haldas:2002}, while others found a transition 
between strongly and weakly localized states but no delocalized 
states. \cite{odagaki:1984, srivastava:1984, koslowski:1990, daboul:2000} More recent work, 
including previous work by this group, has found that a localized-delocalized phase transition 
does in fact exist for the quantum percolation model at finite disorder in two 
dimensions. \cite{islam:2008, schubert:2008, schubert:2009, gong:2009, nazareno:2002}  
Most recently, present authors determined a detailed phase diagram showing a delocalized phase 
at low dilution for all energies $0 < E \leq 1.6$, with a weak power-law localized state 
at higher dilutions and an exponentially localized state at still higher 
dilutions.\cite{dillon:2014} 

Among the various calculations employed to study the quantum percolation model, it stands out 
that most works based on two-dimensional, highly anisotropic strips yield results supporting 
one-parameter scaling's prediction of only localized states, whereas our calculations 
in Ref.~\onlinecite{dillon:2014} and most others finding a delocalized state were based on 
an isotropic square geometry. One of the studies in the first group was by Soukoulis and Grest, 
who used the transfer matrix method to determine the localization length ${\lambda}_{M}$ of 
long, thin, quasi-one-dimensional strips of varying width $M$, after which they used finite size 
scaling to determine the localization length ${\lambda}$ in the two-dimensional limit and thus 
the phase of the system (Ref.~\onlinecite{soukoulis:1991}, see Ref.~\onlinecite{soukoulis:1982} 
for more detail on the transfer matrix method in two dimensions). It is worth noting that 
Daboul et al., who found at most weakly localized states on {\it isotropic} lattices, also found 
disagreement with Soukoulis and Grest's results; they noted that the extreme anisotropy of the 
strips used in the transfer-matrix method might overly influence its results toward a more 1-D 
geometry than 2-D.\cite{daboul:2000} Aside from geometry, Soukoulis and Grest also differ from 
our previous work in that they only examined dilutions within the range $0.15 \leq q \leq 0.50$, 
the lower limit of which is very close to the delocalization phase boundary we found 
in Ref.~\onlinecite{dillon:2014}, which could explain why they did not find any delocalized 
states.  To better understand the differences between our results and theirs, in this work we 
apply our direct calculation of the transmission coefficient to the quasi-1D scaling geometry 
used by Soukoulis and Grest over the same energies $E$ and widths $M$ they used, but over a 
larger range of dilutions extending into lower dilutions than those they examined. We 
additionally examine the inverse participation ratio of the lattices, which, when extrapolated to 
the thermodynamic limit, provides another indicator of localization.

We start from the approach used by Dillon and Nakanishi \cite{dillon:2014} using the quantum 
percolation Hamiltonian with off-diagonal disorder and zero on-site energy 

\begin{equation}
H = \sum_{<ij>} V_{ij}|i\rangle \langle j| + h.c
\label{eq1}
\end{equation}

\noindent where $|i>$ and $|j>$ are tight binding basis functions and $V_{ij}$ is a binary 
hopping matrix element between sites $i$ and $j$ which equals a finite constant $V_{0}$ if 
both sites are available and nearest neighbors and equals 0 otherwise. As in previous works, 
we normalize the system energy and use $V_{0} =1$.

We realize this model on an anisotropic square lattice of varying widths $M$ and lengths $N$ 
to which we attach semi-infinite input and output leads at diagonally opposite corners and 
which we randomly dilute by removing some fraction $q$ of the sites, thus setting their 
corresponding $V_{ij}$ to zero. $N$ is chosen such that $N=10*M$ at minimum to obtain 
quasi-one-dimensional geometry, and such that the diagonally opposite sites have the same parity 
in order to maintain the symmetry of the input and output leads 
as $N$ is varied for the same $M$ (i.e., the leads are always on the same sublattice). The wavefunction for the entire lattice plus input and output leads can be calculated by solving the Schr\"{o}dinger equation

\begin{equation}
\begin{array}{l}
H\psi = E\psi  \\
\mbox{where}, \psi = \left[ \begin{array}{c} \vec{\psi}_{in} \\ 
                     \vec{\psi}_{cluster} \\ \vec{\psi}_{out} \end{array}
               \right]
\end{array}
\label{eq2}
\end{equation}

\noindent and $\vec{\psi}_{in} = \left\{\psi_{-(n+1)}\right\}$ and 
$\vec{\psi}_{out} = \left\{\psi_{+(n+1)}\right\}$,
$n = 0,1,2 \ldots$, are the input and output lead parts of the wave function respectively. 
Using an ansatz by Daboul et al \cite{daboul:2000}, we assume that the input and output parts 
of the wavefunction are plane waves

\begin{equation}
\begin{array}{l}
\psi_{in} {\rightarrow} \psi_{-(n+1)} = e^{-in\kappa} + re^{in\kappa} \\
\psi_{out} {\rightarrow} \psi_{+(n+1)} = te^{in\kappa}
\end{array}
\label{eq3}
\end{equation}

\noindent where $r$ is the amplitude of the reflected wave and $t$ is the amplitude of the 
transmitted wave. This ansatz reduces the infinite-sized problem to a finite one including only 
the main $M \times N$ lattice and the nearest input and output lead sites, for the wavevectors 
${\kappa}$ that are related to the energy $E$ by:

\begin{equation}
E = e^{-i\kappa} + e^{i\kappa}
\label{eq4}
\end{equation}

Note that the plane-wave energies are thereby restricted to the one-dimensional range of 
$-2 \leq E \leq 2$ rather than the full two-dimensional energy range $-4 \leq E \leq 4$; 
nonetheless this energy range has been sufficient for us to observe the localization behavior 
of the wavefunction in prior work. 

\begin{figure}[tb]
{\resizebox{1.5in}{!}{\includegraphics[trim=10 10 450 610]{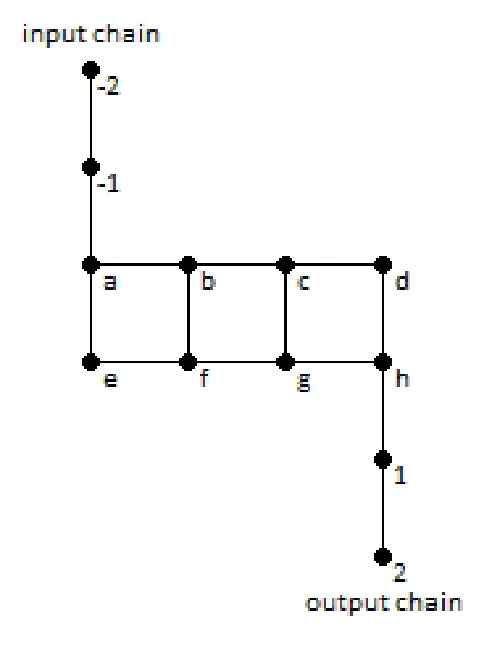}}}
\caption{A toy example of an anisotropic lattice with input and output leads attached to 
opposite corners.}
\label{lattice}
\end{figure}

The reduced Schr\"{o}dinger equation after applying the ansatz can be written as a 
$(M*N +2)\times(M*N +2)$ matrix equation of the form

\begin{equation}
\left[ \begin{array}{ccc}
         -E + e^{i\kappa}  & \vec{c_1}^t  &   0 \\
               \vec{c_1} & \begin{array}[t]{c}
                             {\bf A}
                         \end{array}              &    \vec{c_2} \\
               0     & \vec{c_2}^t  &  -E + e^{i\kappa}
          \end{array} \right]
          \left[ \begin{array}{c}
                    1 + r \\ \vec{\psi}_{clust} \\ t
                    \end{array} \right]
         = \left[ \begin{array}{c}
                  e^{i\kappa} - e^{-i\kappa} \\ \vec{0} \\ 0
                  \end{array} \right]
\label{eq5}
\end{equation}

\noindent where ${\bf A}$ is an $M*N \times M*N$  matrix representing the connectivity of the 
cluster (with $–E$ as its diagonal components), $\vec{c_i}$ is the $M*N$ component 
vector representing the coupling of the leads to the cluster sites, and $\vec{\psi}_{clust}$ 
and $\vec{0}$ are also $M*N$ component vectors, the former representing the wavefunction 
solutions (e.g. on sites $a$-$h$ in Fig. ~\ref{lattice}). The cluster connectivity in 
${\bf A}$ is represented with $V_{ij}=1$ in positions $A_{ij}$ and $A_{ji}$ if $i$ and $j$ are 
connected, otherwise $0$.  

Eq. ~\ref{eq5} is the exact expression for a 2D system connected to semi-infinite chains with 
continuous eigenvalues within the range $-2 \leq E \leq 2$ as discussed above. The transmission 
($T$) and reflection ($R$) coefficients are determined by $T=|t|^{2}$ and $R=|r|^{2}$. 

From the wavefunction solutions of Eq.~\ref{eq5} we also calculate the Inverse Participation 
Ratio (IPR), which measures the fractional size of the particle wavefunction across the lattice 
and gives a picture of the transport complementary to the picture provided by the transmission 
coefficient alone. The IPR is defined here by: 

\begin{equation}
IPR = \frac{1}{\sum_{i} |\psi_i|^{4} (M*N)}
\label{eq6}
\end{equation}

\noindent where ${\psi_i}$ is the amplitude of the normalized wavefunction for the main-cluster 
portion of the lattice on site $i$ and $M*N$ is the size of the lattice. (For our model, we have 
chosen to normalize the IPR by the lattice size rather than connected cluster size, as is 
sometimes done, since the lattice size is the fixed parameter and doing so allows better 
comparison between different sizes when extrapolating to the thermodynamic limit.) It should 
be noted that our $\vec{\psi}$ for given $E$ is a continuum eigenstate of the system containing 
the 1D lead chains, and $\vec{\psi}_{clust}$ is expected to correspond to a mixed state 
consisting of eigenstates of the middle square portion of the lattice. We see that given two 
lattices of the same size, the one with the smaller IPR has the particle wavefunction residing 
on a smaller number of sites, though the precise geometric distribution cannot be known from 
the IPR alone. The IPR is often used to assess localization by extrapolating it to the 
thermodynamic limit; if the IPR approaches a constant fraction of the entire lattice, there are 
extended states, whereas if it decays to zero the states are localized.

The remainder of this paper is organized as follows. In Section II, we study the transmission 
coeffient curves on the highly anisotropic lattices, scaling first to the quasi-one-dimensional 
limit and then to the two-dimensional limit, and find a delocalization-localization transition 
consistent with our previous results. In Section III, we examine the inverse participation ratio 
as a function of lattice size, and find that the IPR's behavior also indicates delocalized 
states being possible. In Section IV, we summarize and analyze our results. 

\section{Transmision coefficient fits}

\begin{figure}[tb]
\vspace{0.4in}
\centering
\resizebox{3in}{!}{\includegraphics[trim=0 0 0 30]{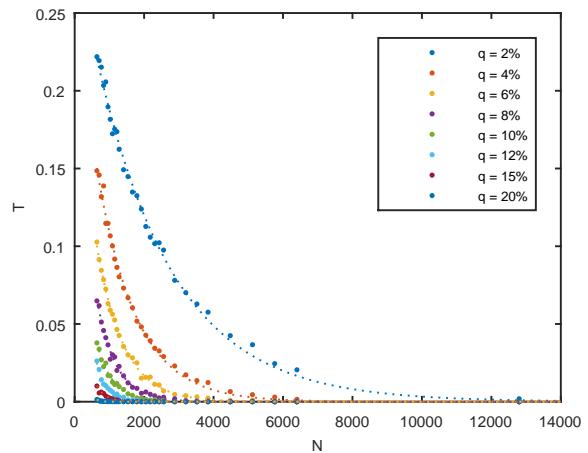}}\\
\caption{Transmission $T$ vs length $N$ for selected dilutions $q$ on lattices with width 
$M = 64$ and particle energy $E = 1.05$. All transmission curves are shown fitted to an 
exponential curve (dotted line).}
\label{TvsN}
\end{figure}

We calculated the transmission $T$ at the energies $E=0.05$, $0.25$, and $1.05$ for anisotropic 
square lattices of width $M=8, 16, 32, 64,$ and $128$ and lengths $N$ within the range 
$10 M \leq N \leq 200 M$, and dilutions $2\% \leq q \leq 50\%$. The energies, widths, and 
dilutions were chosen to match those studied by Ref \onlinecite{soukoulis:1991}, though 
additional dilutions $q < 15\%$ were incorporated since our previous work showed a delocalized 
region at low dilution. The values of $N$ were chosen as $a*M$, where $a$ is an arbitrary/odd integer for even/odd $M$, so that $N$ always has the same parity as $M$ and the input and output leads connected to the corners are always on the same sublattice of the bipartite lattice as $N$ is increased. This ensures that the transport symmetry is not disturbed as $N$ is changed for the given $M$. In our case only even widths $M$ were examined. The upper limit of $N$ was determined by computational limitations, since 
for large $N$ and $q$ the transmission was small enough to result in an underflow 0. For most 
dilutions this occurred for $N \geq 200 M$, though for larger energies the cut-off was lower. 
Despite these computational limitations in calculating $T$, we found that the transmission 
dropped off with $N$ sufficiently smoothly and quickly that an accurate fit of the transmission 
was able to be found for all but the highest dilutions ($q \geq 30\%$), all of which fall well 
above the delocalization-localization phase boundary found in Ref. \onlinecite{dillon:2014}.


From the transmission coefficients for lattices with the above parameters we are able to 
determine the localization length $\lambda_{M}$  of the various strips of width $M$, and 
extrapolate the results in the isotropic limit to find the 2D localization length $\lambda$. 
First, we plot transmission $T$ vs lattice length $N$ for each width $M$ and energy $E$, for 
each dilution $q$ that had a sufficient number of points to establish a fit for the transmission 
($2\% \leq q \leq 25\%$ for smaller $M$ and larger $E$). All dilution curves decay exponentially 
($T=a*exp(-b N)$), as is to be expected given the highly anisotropic quasi-1D lattices. 
An example of the fitted $T$ vs $N$ curves for $E=1.05$ and $M=64$ at selected dilutions is 
given in Fig.~\ref{TvsN}.  The other ($E$,$M$) pairs have similar transmission curves. 

\begin{figure}[tb]
\centering
\resizebox{3in}{!}{\includegraphics[trim=0 0 0 0]{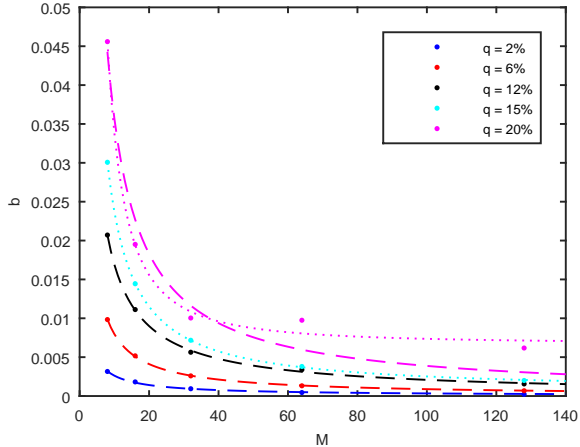}}\\
\caption{Inverse localization length $b_M$ of quasi-1D strips of width $M$ plotted vs. $M$ for 
selected dilutions $q$ at an energy of $E=1.05$. At low dilutions the best fit is one which 
decays to zero, indicating delocalization (dashed line); for higher dilution the best fit is 
one with an offset term corresponding to the inverse localization length of the systerm 
(dotted line).}
\label{Bfits}
\end{figure}

To determine the inverse localizaion length $b_M \propto 1/\lambda_{M}$ in the 
quasi-one-dimensional thermodynamic limit of a strip of width $M$ and length 
$N \rightarrow {\infty}$, we use the successive fitting procedure described in 
Ref. \onlinecite{dillon:2014} Section III. For each $T$ vs $N$ curve, we fit just the first 6 
points, then the first 7 points, etc until all points have been added, saving the parameter $b$ 
from each exponential fit. We then plot the saved $b$ vs $N_{max}$, where $N_{max}$ is the 
maximum length included in the fit resulting in that value of $b$, and fit this new curve to 
find the non-zero value $b_M$ that the curve stablizes to as $N_{max} \rightarrow {\infty}$. 

Extrapolating $b_M$ to the 2D isotropic limit determines the localization of the system: 
for $b_{M} \rightarrow 0$ ($\lambda_{M} \rightarrow {\infty}$) as $M \rightarrow {\infty}$, 
the system is delocalized, but if $b_{M} \rightarrow b_{\infty}$ 
($\lambda_{M} \rightarrow \lambda$), where $b_{\infty}$ is some finite constant,  
then the system is localized. We fit $b_{M}$ vs $M$ for each $E$ and $q$ and find that for 
$E=1.05$ and $q \leq 12\%$, $E=0.25$ and $q \leq 15\%$, and $E=0.05$ and $q \leq 8\%$ , a fit 
which decays to zero is the best fit, whereas above these dilutions a fit with a constant offset 
$b_{\infty}$ fits the $b_M$ vs $M$ curves better, as shown for example at $E=1.05$ 
in Fig.~\ref{Bfits}. Thus there {\it is} in fact a delocalized phase at each energy for these 
low values of disorder. Moreover, while our prior work did not study these energies specifically, 
the upper bounds for delocalization found here correspond roughly to those found in 
Ref. ~\onlinecite{dillon:2014}; an exact match is not expected due to the different geometry. 
At the dilutions and energies for which $b_M$ stabilizes to a non-zero value $b_{\infty}$, 
we calculate the localization length by $\lambda = 1/b_{\infty}$.

\begin{figure}[tbp]
\centering
\resizebox{3in}{!}{\includegraphics[trim=0 0 0 0]{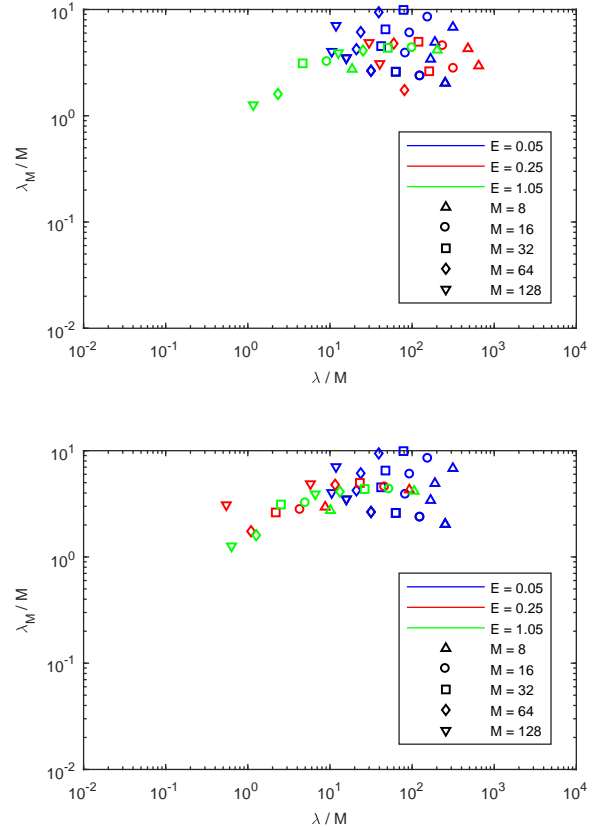}}\\
\caption{Plot of scaled localization lengths $\lambda_{M}/M$ vs $\lambda/M$ for the widths $M$ 
(differentiated by marker type) and energies $E$ (differentiated by color) studied. In (a) we 
use the $\lambda$ extrapolated from the $\lambda_M$ vs $M$ fits, and in (b) we use different 
values for $\lambda$ at $E= 1.05$ and $E = 0.25$ that fall within the determined error bounds 
of the extrapolated value.}
\label{ScaledLambda}
\end{figure}

To most easily compare our results with those of Soukoulis and Grest, we plot $\lambda_{M}/M$ vs 
$\lambda/M$ (see Fig.~\ref{ScaledLambda}a) as in their Fig. 1 from 
Ref. \onlinecite{soukoulis:1991}. There are two noticeable differences and one important 
similarity between their results and ours. First, our figure has no points in the lower left 
for the smaller values of $\lambda$ and $\lambda_M$. This area is where the small localization 
lengths at high dilution should be, and their absence is simply a result of the computational 
limitations of our transmission fit technique, described at the beginning of this section. 
Secondly, the localization lengths we do have do not exactly overlap in their values with those 
of Soukoulis and Grest, nor do they all collapse neatly onto one curve. These might partially be 
explained by a missing factor in our transmission fits; when determining the localization length,
we assumed that in the fit $T=a*exp(-b_M N)$, $b_M = 1/\lambda_M$, when it may be that 
$b_M = c/\lambda_M$, where $c$ is another constant which may depend on $M$. If there is such a 
constant, the vertical position of our localization points may be shifted from their true values.
Additionally, the $\lambda$ determined by Soukoulis and Grest were a fitting parameter chosen to 
induce the points to collapse onto one curve, following the scaling procedure outlined in 
Ref.s~\onlinecite{mackinnon:1981} and ~\onlinecite{soukoulis:1982}, whereas our values of 
$\lambda_{\infty}$ were determined independently. Our $\lambda$ do have error bars (omitted 
from Fig.~\ref{ScaledLambda} to avoid cluttering the figure) and choosing different $\lambda$ 
within the bounds of our fit estimates at $E=1.05$ and $E = 0.25$, for instance, yeilds a 
somewhat better collapse of those energies' localization lengths onto one curve 
(see Fig.~\ref{ScaledLambda}b).

Despite the differences in our two figures, however, there is one significant similarity. 
That is, in the dilution range $15\% \leq q \leq 20\%$ for which our work {\it does} have 
overlap with the dilutions studied by Soukoulis and Grest, our localization lengths fall within 
the same order of magnitude of those found by Soukoulis and Grest. They are not precisely the 
same, but they are not wildly different, either. This gives us confidence that our technique is 
yielding the same results as theirs, leading us to believe that they simply did not look at 
small enough dilutions to see a transition, relying instead on an extrapolation that may not 
be justified. 

\section{Inverse Participation Ratio calculations}

\begin{figure}[tbp]
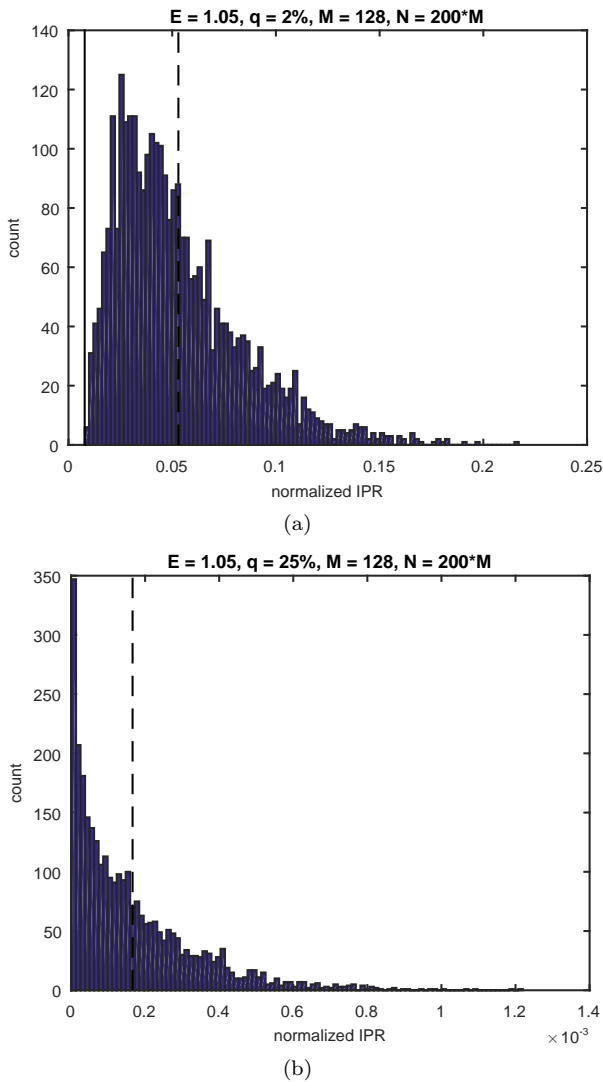

\centering
    \subfigure[]{\label{fig:ipr_histograma} \resizebox{!}{2.6in}{\includegraphics{fig5a.eps}}}
    \subfigure[]{\label{fig:ipr_histogramb} \resizebox{!}{2.6in}{\includegraphics{fig5b.eps}}}
\caption{Examples of the histogram of IPR values calculated at energy $E = 1.05 $ on a 
128x25600 lattice for (a) $q=2\%$ and (b) $q=25\%$.The vertical dashed line on each histogram 
indicates the location of the average IPR for that distribution. For (a), the solid vertical 
line marks 1/M, which is approximately the lower limit for the IPR required to span the lattice; 
it is not shown on (b) because it is exceeds the right hand bound of the figure.}
\label{ipr_histogram}
\end{figure}
\begin{figure}[tb]
\centering
\resizebox{3in}{!}{\includegraphics[trim=0 0 0 0]{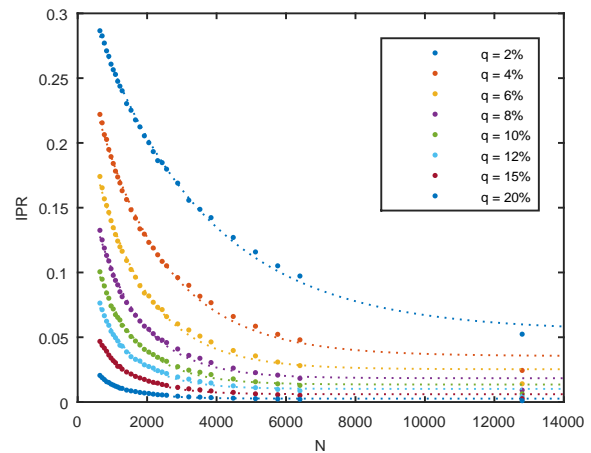}}\\
\caption{Average inverse participation ratio $IPR$ vs length $N$ for selected dilutions $q$ on 
lattices with width $M = 64$ and particle energy $E = 1.05$, shown fitted to a curve with an 
offset (dotted line).}
\label{IPRvsN}
\end{figure}
\begin{figure*}[tb]
\centering
\resizebox{!}{1.5in}{\includegraphics[trim=0 0 0 0]{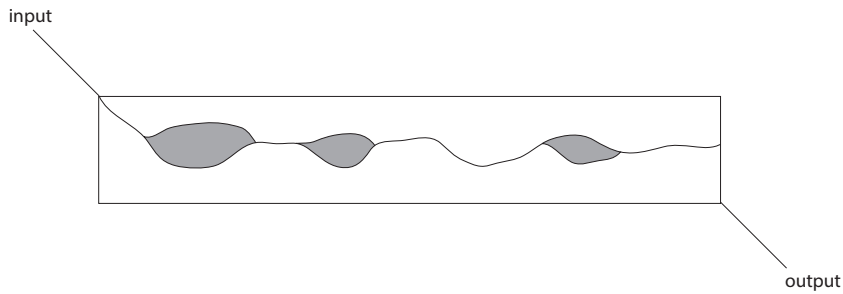}}\\
\caption{An example of a lattice with a connected cluster spanning the lattice but not attached 
to the output corner.}
\label{disconnected}
\end{figure*}

To corroborate our finding delocalized states at small disorder on the anisotropic quasi-1D 
strips, we also examined the Inverse Participation Ratio (IPR) as the system size increases. 
We observe that even at the largest and most anisotropic lattice studied, $M=128, N=200M$, 
the IPR distribution of all realizations at small dilution has a distinct peak at $IPR \neq 0$, 
and the average value is greater than $1/M$, which is the minimum fraction required to span the 
lattice, whereas at large dilution the peak is near zero (Fig.~\ref{ipr_histogram}). This seems 
to hint at the transition we observed in the previous section. When we plot the average IPR 
vs N for fixed width (as in Fig.~\ref{IPRvsN}), we also see that the average IPR decreases 
much less rapidly than the corresponding transmission T (compare Fig.~\ref{TvsN}), with the IPR 
at large N remaining well above zero (in fact, the average IPR does not have the problem with 
computational underflow that the transmission does at large dilution). This illustrates the fact 
that while the transmission and IPR are related, they do represent different ways of examining 
localization. It is entirely possible to have many realizations with clusters spanning the 
lattice that are connected to the input site and to an edge site on the opposite end that is 
not the output site, as illustrated in the hypothetical example in Fig.~\ref{disconnected}. 
If this is the case, the average IPR (which is measured over the entire lattice irrespective 
of the input) would be nonzero while the transmission (which is measured only corner-to-corner) 
would be very close to zero. 

\begin{figure}[tb]
\centering
\resizebox{3in}{!}{\includegraphics[trim=0 0 0 0]{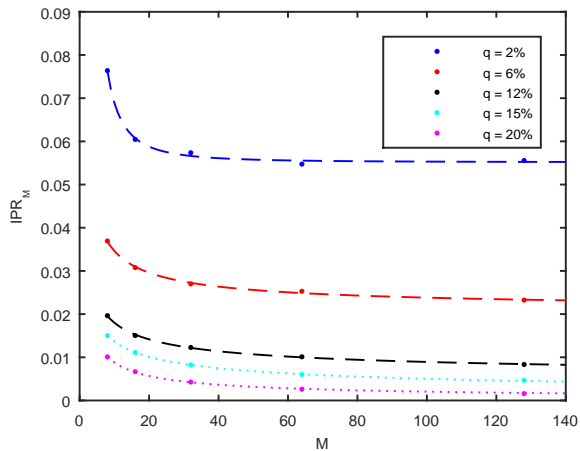}}\\
\caption{Inverse participation ratio $IPR_M$ for fixed-width lattices plotted vs lattice width 
$M$ for selected dilutions $q$ at energy $E = 1.05$. For low dilution, 
$IPR_M \rightarrow IPR_{\infty}$ as $M \rightarrow {\infty}$ (dashed lines), indicating 
delocalization, while at higher dilutions $IPR_M \rightarrow 0$, indicating localization 
(dotted line)}
\label{IPRfits}
\end{figure}

When we fit the average $IPR$ vs $N$ curves at fixed $M$, we observe an interesting trend in the 
curves as dilution increases. Surprisingly, the $IPR$ vs $N$ at low dilution can be fit very well
to a curve with a nonzero offset that we call $IPR_M$. An example at $E=1.05$ and $M=64$ is shown 
in Fig.~\ref{IPRvsN} This is unexpected, since in the 1-D limit we know the states are localized, 
but can be explained as being a result of our including lengths only up to $N=200M$ due to the 
computational limitations in $T$. However, as $q$ increases to large disorder, we find that IPR 
vs $N$ is best fit by a curve decreasing smoothly to zero. This change hints at the phase 
transition we observed in the 2D isotropic limit in Section II: if we included still longer 
lengths N toward the 1-D limit, we should see the $IPR \rightarrow 0$ since the 1D limit is 
purely localized, but if we extrapolate toward an isotropic case, we should see the average IPR 
approach a finite value for those dilutions at which we found delocalization. To capture the 
latter situation, we plot the offset terms $IPR_M$ from the lower dilution fits vs $M$ 
(see Fig.~\ref{IPRfits} for example at $E = 1.05$). Again, we find that at small dilution the 
$IPR_M$ vs $M$ curve is best fit by a curve with an offset (in this case, a power-law with 
offset), meaning the IPR grows in proportion to the width and stabilizes to a nonzero fraction 
of the lattice as we scale toward the isotropic 2D limit. On the other hand, as we increase 
the dilution, the $IPR_M$ vs $M$ curves eventually are better fit by a pure power law, meaning 
that although the anisotropic lattices may have had spanning clusters, these clusters do not 
grow proportionally with $M$ and eventually become disconnected from the output edge, resulting 
in localization. For $E = 1.05$ this shift to pure power-law fit occurs at $q \geq 15\%$, for 
$E=0.25$ at $q \geq 15\%$, and for $E = 0.05$ at $q \geq 8\%$. The results of the IPR study 
demonstrate that there are spanning clusters in the isotropic limit at low dilution, meaning 
there are indeed delocalized states at these dilutions, with a transition to a localized state 
(isolated clusters) at sufficiently high disorder. Moreover, the transition to localized states 
as disorder increases occurs at or very near the same dilutions at which we found a transition 
using the transmission calculations.

\section{Summary and Conclusions}

We have studied the quantum percolation model on highly anistotropic two-dimensional lattices, 
scaling toward the isotropic two-dimensional case (studied in previous works) to determine the 
localization state and localization length in the thermodynamic limit. We determined the 
localization length by a two-step process in which we first determined the inverse localiztion 
length $b_M = \lambda_M^{-1}$ of the anisotropic strips by extrapolating $N \rightarrow \infty$, 
then extrapolated $\lambda_M$ to the localization length $\lambda$ of the isotropic system from 
the trend of the $b_M$ as $M \rightarrow {\infty}$. Although the transmission calculations only 
allow us to study a limited range of dilutions effectively due to computational limitations, 
we nonetheless were able to detect a phase transition at specific dilutions, above which the 
$M$-width strip inverse localization lengths $b_M$ converged to a finite value, and below which 
they decayed to zero, indicating an infinite, lattice-spanning extended state. The location of 
the phase transitions are consistent with the phase boundaries found in our previous work 
(Ref. ~\onlinecite{dillon:2014}), but their existence is in contradition to the results 
predicted, e.g., by Soukoulis and Grest in their transfer-matrix method studies of quantum 
percolation. \cite{soukoulis:1991} This contradiction can be resolved by observing that they 
only studied dilutions above $q = 15\%$, which is above the delocalization-localization phase 
boundary found in this work and our prior work. The localization lengths found in this work for 
dilutions within the localized region fall within the same order of magnitude of those found 
by Soukoulis and Grest at the lower end of the range of dilutions they studied, leading us to 
believe that they simply did not look at small enough dilutions, thus missing the phase 
transition.

We additionally checked the localization state of the anisotropic strips by studying the inverse 
participation ratio of the lattices, which tells us what fraction of sites sustain the particle 
wavefunction. We find that even on narrow anisotropic strips, at small dilution the average 
IPR shows a distinct peak away from zero at a value large enough to span the lattice, while at 
large dilutions the peak is near zero. When we scale toward the isotropic limit, we find that 
the IPR vanishes for large dilution, indicative of localization, while it approaches a finite 
value for low dilution, indicative of a delocalized state. Furthermore, the dilutions above 
which the inverse participation ratio vanishes in the isotropic limit match the phase boundaries 
found in the first part of the paper. 

The results of our work in this paper serve two purposes. First, by using the same basic 
technique (transmission coefficient and inverse participation ratio measurements) as our 
previous work on a different geometry - that is, highly anisotropic lattices scaled to the 2D 
thermodynamic limit - we obtain the same delocalization-localization phase boundary results, 
showing that the phase transition found previously was not dependent on using isotropic geometry. 
Secondly, by using the same geometry as Soukoulis and Grest (and other works that use the
transfer matrix approach), we found overlap between our 
localization length results and theirs at higher dilutions, but also examined smaller dilutions 
and found a delocalized state, leading us to suspect that their extrapolation of the results 
from $15\% \leq q \leq 50\%$ toward even smaller dilutions was not warranted. Had our 
localization lengths within their range dramatically differed from theirs, we would perhaps 
conclude that the differing techniques used led to the difference in whether a delocalized state 
was found, but as we have shown, this seems not to be the case.

\section*{Ackowledgements}

We thank Purdue Research Foundation and Purdue University Department of Physics for financial
support and the latter for generously providing computing resources.

\end{document}